\documentclass[twocolumn,english,nobibnotes,nofootinbib,superscriptaddress ]{revtex4-1}
\usepackage[T1]{fontenc}
\usepackage[latin9]{inputenc}
\usepackage{geometry}
\usepackage{amsbsy}
\usepackage{graphicx}
\usepackage{babel}
\usepackage{color}
\begin{document}

\title{Electronic structure of (1e,1h) states of carbon nanotube quantum dots}

\author{E. N. Osika}
\affiliation{AGH University of Science and Technology, Faculty of Physics and Applied Computer Science,\\
al. Mickiewicza 30, 30-059 Krak\'ow, Poland}
\affiliation{ICFO-Institut de Ci\'encies Fot\'oniques, The Barcelona Institute of Science and Technology, 08860 Castelldefels (Barcelona), Spain
}

\author{B. Szafran}
\affiliation{AGH University of Science and Technology, Faculty of Physics and Applied Computer Science,\\
al. Mickiewicza 30, 30-059 Krak\'ow, Poland}

\date{\today}

\begin{abstract}
We provide an atomistic tight-binding description of a few carriers confined in ambipolar ($n$-$p$) double quantum 
dots defined in a semiconducting carbon nanotube. We focus our attention on the charge state of the system 
 in which Pauli blockade of the current flow is observed [F. Pei {\it et al.}, Nat. Nanotechnol.  {\bf 7}, 630 (2012);
E. A. Laird {\it et al.}, {\it ibid}  {\bf 8}, 565 (2013)] with a single excess electron in the $n$-dot and a single hole in the $p$-dot. We use the configuration interaction approach to determine the spin-valley structure of the states near the neutrality point and discuss its consequences for the interdot exchange interaction, the degeneracy of the energy spectrum and the symmetry of the confined states. We calculate the transition energies lifting the Pauli blockade and analyze their dependence on the magnetic field vector. 
 Furthermore, we introduce bending of the nanotube and demonstrate its influence on the transition energy spectra.
The best qualitative agreement with the experimental data is observed for nanotubes deflected in the gated areas in which the carrier confinement is induced.
\end{abstract}

\maketitle
\section{Introduction}

Semiconducting carbon nanotubes (CNTs) \cite{cnt} provide a clean medium for carrier confinement and confined spin manipulation by electric fields. The latter is made possible due to the spin-orbit interaction \cite{Ando,St,Dh,Jk,Lc,Md,Bu,Iz} that translates the electron motion in space to the rotation of its spin. The spin-orbit coupling in CNTs is strong \cite{St} and originates  from activation of the atomic spin-orbit coupling accompanying the folding of the graphene plane into a tube \cite{rmp}. The spin-orbit interaction governs
 the spin-valley structure of the states near the charge neutrality point in both the conductance and the valence bands \cite{kuemmeth} forming spin-valley doublets separated by an energy gap
known as zero-field splitting \cite{St}.  The doublets that can serve as qubits, the absence of the dephasing \cite{dephas} due to the nuclear spin field, and the strong spin-orbit coupling  make the CNTs interesting for quantum information storage and processing  \cite{zutic,hanson,loss}. 

The spin transitions in quantum dots (QDs) are monitored experimentally via the spin-valley blockade \cite{buitelaar,lairdpei,peilaird,palyib,papa} of the current flow across the double dot system. 
The current blockade is lifted by the electric dipole spin resonance (EDSR) driven by ac external voltages \cite{lairdpei,peilaird,osimre,sz,wa,bent,lilaird}
for resonant frequencies. A large amount of theoretical work was performed for the 
description of the charge configuration with a single electron per dot (1e,1e). In particular, the spin-valley blockade in the presence of a short-range disorder was discussed \cite{palyib}.
The symmetries of the confined states in the presence of the spin-orbit coupling were analyzed \cite{stecher,weiss}. The spin dephasing due to the hyperfine interaction with $^{13}$C atoms 
was discussed for charge dynamics in a (0,2e)$\leftrightarrow$(1e,1e) cycle \cite{rf}. The leakage current across Pauli-blocked double dots in straight \cite{palyi_straight} and bent \cite{palyi_shape}  CNT was also studied. Nevertheless, in CNTs the EDSR was observed only for ambipolar double dots \cite{lairdpei,peilaird}, with a single electron occupying a $n$-type dot and a single hole stored in the $p$-type dot. In the present work this charge configuration is denoted as (1e,1h). 
The EDSR transition spectra  were recently analyzed \cite{lilaird} in a continuum description that produced good agreement with the experiment for model parameters with realistic values. 
To the best of our knowledge Ref. \cite{lilaird} was the only theoretical paper to consider the (1e,1h) system in CNT double dots. In (1e,1h) the spin-valley structures of the 
$n$ and $p$ dots are not identical. Moreover,  a strong tunnel coupling between the dots is difficult to obtain, since for the coupling to appear the conduction-band electron of the $n$ dot has to climb the potential maximum forming the $p$ dot \cite{weaker}. The tunnel coupling in turn is necessary for the exchange \cite{exchange} interaction to appear.

\begin{figure*}[htbp]
\includegraphics[scale=0.4]{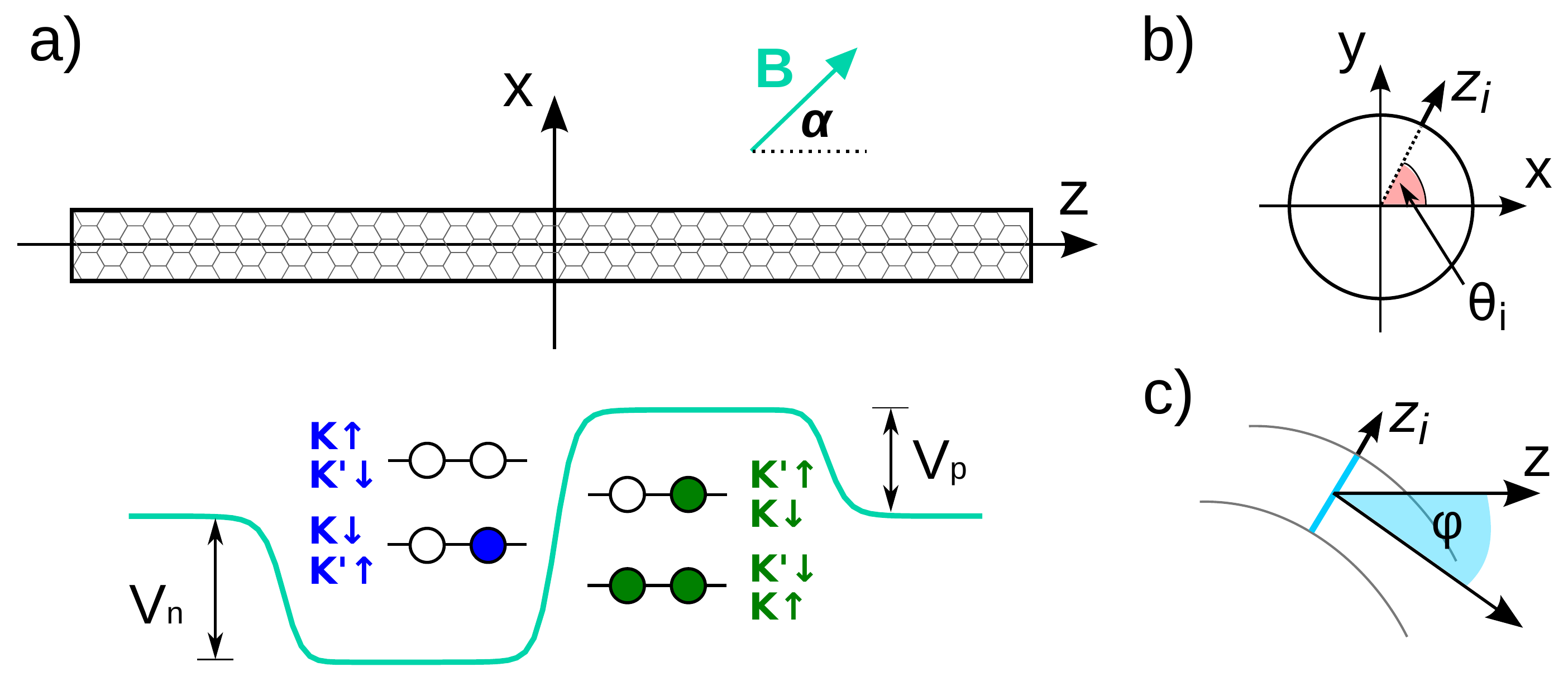}
\caption{(a) Scheme of the considered system - CNT with a QD of depth $V_n$ and an antidot of height $V_p$.
The $n$ dot contains one electron at the bottom of the conduction band and the $p$ dot contains three electrons
at the top of the valence band (a single confined hole).
The lower plot shows the profile of the external potential, the considered charge configuration {[}denoted as (1e,1h) in the text{]} and the degeneracy of single-electron energy levels at $B=0$. 
The magnetic field vector is applied within the $xz$ plane  and forms an angle $\alpha$ with the $z$ axis.
(b) Azimuthal angle $\theta$ and (c) an angle $\phi$ of a local bent of the CNT. The CNT is considered bent within the $xz$ plane.  }\label{schemat}
\end{figure*}

In this paper we determine the electronic structure of the (1e,1h) charge configuration of the double QD using the tight-binding approach and the exact diagonalization method for the problem of a few carriers localized in states near the neutrality point. We account for the spin-orbit interaction of the orbital type. 
The present  approach due to its atomistic nature 
describes all the intervalley scattering effects \cite{rontani} that result from the electron-electron interaction.
The tunnel coupling between the carriers of $n$- and $p$-type QDs as well as the deformations of the CNT axis are also accounted for at the atomic level.
We  discuss the spin-valley symmetry of states in the low-energy part of the spectrum and analyze the effects of the exchange interaction for the electronic structure near the ground-state of the (1e,1h) system. We calculate the energies of transitions that unblock the Pauli
blockade and demonstrate their essential dependence on the deflection of the CNT axis.

A bend in a CNT translates the strongly anisotropic effective Land\'e $g$ factor 
into a position dependent effective magnetic field which was indicated earlier \cite{bent} as a prerequisite for driving the spin transition by EDSR.
The original model \cite{bent} of the bend and the subsequent ones \cite{lilaird,palyi_shape} assumed that the deflection of the CNT does not change along each of the QDs.
However, in the experiments the CNT axis is bent due to its deposition above the metal gates \cite{lairdpei}. The model applied in this paper accounts for corrections to the spin-orbit interaction stemming from the local deflection of the CNT axis \cite{OsikaJPCM}. The spin-orbit coupling sets the spins of the eigenstates near the neutrality point aligned parallel or antiparallel to the axis of the CNT. 
We study the consequences of the bent axis on the resonant transition energies lifting the Pauli blockade, in particular for their dependence on the orientation of the external magnetic field. The details of the bend drastically change the dependence of the transition energies for  the magnetic field rotated off the CNT axis.
Only when the bent parts of the CNT are above the metallic gates inducing the QD potentials
a good qualitative agreement of the calculated transition spectrum to the experiment data is obtained.

This paper is organized as follows. The next section contains the theory. The results and discussion
are given in Sec. III. The straight CNTs are considered in Sec. III A with the single-particle spectra analyzed in Sec. III A 1, the description of the few-carrier states in Sec. III A 2, and the transitions spectra in Sec. III A 3. The results for bent CNTs are given in Sec. III B.
The Summary and Conclusions are provided in Sec. IV.


\section{Theory}
We consider a semiconducting nanotube of zigzag chirality, length $L=53.11$ nm, and diameter
$2r=1.33$ nm. The considered nanotube 
 has 17 atoms along the circumference with the chiral vector $C_{h}=(17,0)$.
 The center of the nanotube is set at $z=0$.
For a straight CNT its axis coincides with the $z$-axis of the adopted reference system {[}cf. Fig. \ref{schemat}(a){]}.
The deflected CNTs are assumed bent within the $xz$-plane {[}cf. Fig. \ref{schemat}(c){]}.
The bend is locally parametrized by the inclination angle $\phi_{i}$ 
of the local CNT axis (for $i$-th ion) to the global $z$ axis {[}cf.
Fig. \ref{schemat}(c){]}.
We model the ambipolar ($n$-$p$) double QD  confinement 
using the formula,
\begin{equation}
W_{QD}(z)=-\frac{V_{n}}{1+(\frac{z+z_{s}}{d})^{14}}+\frac{V_{p}}{1+(\frac{z-z_{s}}{d})^{14}},
\end{equation}
where $V_{n}$ ($V_{p}$ ) is a depth (height) of the $n$-type ($p$-type) dot
as defined by  gate electrodes,
$2d$ stands for the width of the single dot and $z_{s}$ is a shift
of the $n$ ($p$) dot to the left (right) from the origin {[}cf. Fig. \ref{schemat}(a){]}.
 The $n$-type QD confines electrons from
the conduction band while the $p$-type dot confines
electrons from the valence band (or holes in the unoccupied states).
In the calculations we use $2d=16$ nm 
and $z_{s}=8$ nm.

In the exact diagonalization method we first calculate  the single-electron states using the tight-binding
approach. For this purpose we solve the eigenequation for the single-electron
Hamiltonian
\begin{eqnarray}
&H_{1e}&=\sum_{\{i,j,\sigma,\sigma'\}}(c_{i\sigma}^{\dagger}t_{ij}^{\sigma\sigma'}c_{j\sigma'}+h.c.)\nonumber \\ &+&\sum_{i,\sigma,\sigma'}c_{i\sigma}^{\dagger}\left(W_{QD}({\bf r}_{i})\delta_{\sigma\sigma'}+\frac{g_L\mu_{b}}{2}\boldsymbol{\sigma}^{\sigma\sigma'}\cdot{\bf B}\right)c_{i\sigma'},\label{ham}
\end{eqnarray}
where the first summation runs over the $p_{z}$ spin-orbitals of the nearest
neighbor pairs of atoms, $c_{i\sigma}^{\dagger}\ensuremath{}(c_{i\sigma})$
is the particle creation (annihilation) operator at ion $i$ with
spin $\sigma$, and $t_{ij}^{\sigma\sigma'}$ is
the spin-dependent hopping parameter. The second summation in Eq. (\ref{ham}) accounts
for the QD potential and the Zeeman interaction due to external
magnetic field ${\bf B}=(B_{x},0,B_{z})$ that is applied within the $xz$ plane.
In Eq. (\ref{ham}), $\delta_{\sigma\sigma'}$ is the Kronecker delta, $g_L=2$ is the Land\'e factor,
$\mu_{b}$ is the Bohr magneton and $\boldsymbol{\sigma}$ stands
for the vector of Pauli matrices.

In the model we account for the spin-orbit interaction due to curvature
of the graphene plane \cite{Ando} including  the bend
of the CNT axis as described in \cite{OsikaJPCM}. The spin-orbit interaction due to curvature mixes the
$p_{z}$ orbitals (labeled $z_{i}$ for $i$-th ion) with $p_{x}$
and $p_{y}$ orbitals of opposite spins. 
Following Refs. \cite{Ando,OsikaJPCM}
 the formulae for the  spin-dependent hopping parameters read

$t_{ij}^{\uparrow\uparrow}=(z_{i}|H|z_{j})+i\delta\cos\phi_{j}(z_{i}|H|x_{j})-i\delta\cos\phi_{i}(x_{i}|H|z_{j})+i\delta\sin\phi_{j}\sin\theta_{j}(z_{i}|H|y_{j})-i\delta\sin\phi_{i}\sin\theta_{i}(y_{i}|H|z_{j}),$

$t_{ij}^{\downarrow\downarrow}=(z_{i}|H|z_{j})-i\delta\cos\phi_{j}(z_{i}|H|x_{j})+i\delta\cos\phi_{i}(x_{i}|H|z_{j})-i\delta\sin\phi_{j}\sin\theta_{j}(z_{i}|H|y_{j})+i\delta\sin\phi_{i}\sin\theta_{i}(y_{i}|H|z_{j}),$

$t_{ij}^{\uparrow\downarrow}=-i\delta\sin\phi_{j}(z_{i}|H|x_{j})+i\delta\sin\phi_{i}(x_{i}|H|z_{j})-\delta(\sin^{2}\frac{\phi_{j}}{2}e^{i\theta_{j}}+\cos^{2}\frac{\phi_{j}}{2}e^{-i\theta_{j}})(z_{i}|H|y_{j})+\delta(\sin^{2}\frac{\phi_{i}}{2}e^{i\theta_{i}}+\cos^{2}\frac{\phi_{i}}{2}e^{-i\theta_{i}})(y_{i}|H|z_{j}),$

and

$t_{ij}^{\downarrow\uparrow}=-i\delta\sin\phi_{j}(z_{i}|H|x_{j})+i\delta\sin\phi_{i}(x_{i}|H|z_{j})+\delta(\sin^{2}\frac{\phi_{j}}{2}e^{-i\theta_{j}}+\cos^{2}\frac{\phi_{j}}{2}e^{i\theta_{j}})(z_{i}|H|y_{j})-\delta(\sin^{2}\frac{\phi_{i}}{2}e^{-i\theta_{i}}+\cos^{2}\frac{\phi_{i}}{2}e^{i\theta_{i}})(y_{i}|H|z_{j}).$

The matrix elements for the neighbor $p_{z}$ orbitals used
above are $(\gamma_{i}|H|\gamma_{j})=V_{pp}^{\pi}{\bf n}(\gamma_{i})\cdot{\bf n}(\gamma_{j})+(V_{pp}^{\sigma}-V_{pp}^{\pi})\frac{({\bf n}(\gamma_{i})\cdot{\bf R}_{ji})({\bf n}(\gamma_{j})\cdot{\bf R}_{ji})}{{\bf |R}_{ji}|^{2}},$
where $\gamma=x,\, y$ or $z$, $\gamma_{i}$ is the orbital 
of ion at ${\bf R}_{i}$ position, and ${\bf n}(\gamma_{i})$ is a
unit vector in the direction of orbital $\gamma_{i}$. Below, we use the spin-orbit coupling parameter
$\delta=0.003$ \cite{Ando,Md} and the tight-binding
parameters $V_{pp}^{\pi}=-2.66$ eV, $V_{pp}^{\sigma}=6.38$ eV \cite{Tomanek}.
The orbital effects of the magnetic field are taken into account by inclusion
of the Peierls phase 
to the hopping parameters 
\[
t_{ij}^{\sigma\sigma'}(B)=t_{ij}^{\sigma\sigma'}(0)\exp(i\frac{2\pi}{\Phi_{0}}\int_{{\bf r}_{i}}^{{\bf r}_{j}}{\bf A}\cdot{\bf dl}),
\]
where $\Phi_{0}=h/e$ is the flux quantum, ${\bf B}=\nabla\times{\bf A}$,
and ${\bf A}=(0,B_{z}x,B_{x}y)$ (the Landau gauge).

For the calculations of the many-electron states the configuration interaction
(CI) method is used with the Hamiltonian
\begin{equation}
H_{4e}=\sum_{a}\epsilon_{a}g_{a}^{\dagger}g_{a}+\frac{1}{2}\sum_{abcd}V_{ab;cd}g_{a}^{\dagger}g_{b}^{\dagger}g_{c}g_{d}, \label{4e}
\end{equation}
where $g_{a}^{\dagger}$ ($g_{a}$) is the electron creation (annihilation)
operator in the eigenstate $a$ of the single-electron Hamiltonian
$H_{1e}$, $\epsilon_{a}$ is the energy level of the state $a$, and $V_{ab;cd}$ are
the electron-electron interaction matrix elements.  The latter are integrated
in the real and spin space, according to formula
\begin{eqnarray}
V_{ab;cd}&=&\langle\psi_{a}({\bf r_{1}},\boldsymbol{\sigma}_{1})\psi_{b}({\bf r_{2}},\boldsymbol{\sigma}_{2})|H_{c}|\psi_{c}({\bf r_{1}},\boldsymbol{\sigma}_{1})\psi_{d}({\bf r_{2}},\boldsymbol{\sigma}_{2})\rangle\nonumber \\
&=&\sum_{i,\sigma_{i};j,\sigma_{j};k,\sigma_{k};l,\sigma_{l}}\beta_{i,\sigma_{i}}^{a*}\beta_{j,\sigma_{j}}^{b*}\beta_{k,\sigma_{k}}^{c}\beta_{l,\sigma_{l}}^{d}\delta_{\sigma_{i};\sigma_{k}}\delta_{\sigma_{j};\sigma_{l}}\times \nonumber\\&&\langle p_{z}^{i}({\bf r}_{1})p_{z}^{j}({\bf r_{2}})|H_{C}|p_{z}^{k}({\bf r_{1}})p_{z}^{l}({\bf r_{2}})\rangle,
\end{eqnarray}
\begin{figure*}[htbp]
\includegraphics[scale=0.28]{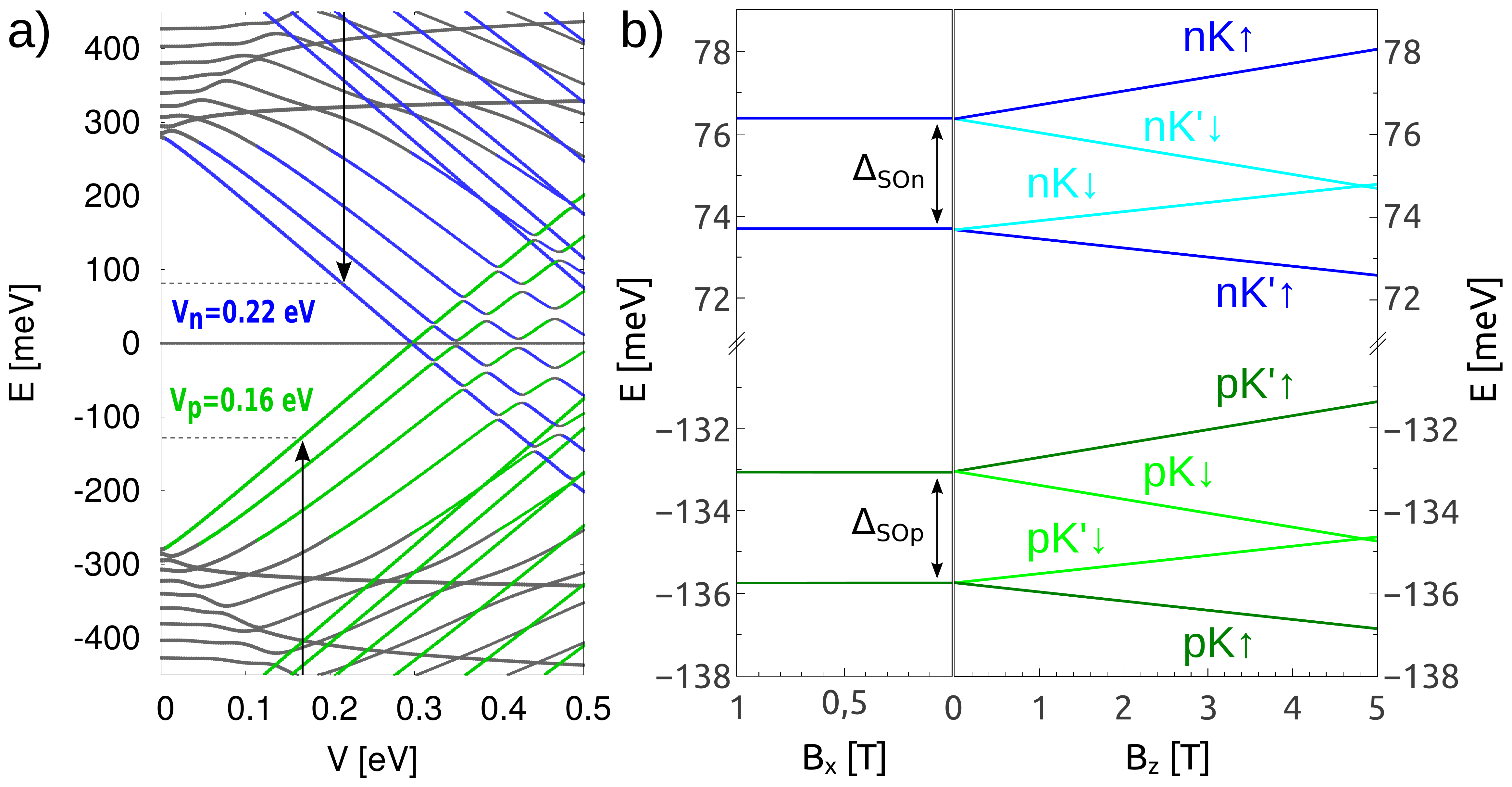}
\caption{(a) Single-electron energy levels as functions of $V=V_{n}=V_{p}$.
The results with the blue/green lines correspond to the majority (at least 75\%) of
the charge density localized within the quantum $n$/$p$ dot. The arrows in (a) indicate the workpoint
chosen below for discussion of (1e,1h) ground-state configuration [see Fig. \ref{schemat}(a)]
with  $V_{n}=0.22$ eV, $V_{p}=0.16$ eV. 
(b) Dependence on external magnetic field $B_{x}$ and $B_z$ for voltages indicated in (a). Captions 
identify the spin ($\uparrow$ or $\downarrow$) 
and valley ($K$ or $K'$) states of the single electron levels.
$\Delta_{SOn}$ and $\Delta_{SOp}$ are the spin-orbit splitting energies.} \label{spectra}
\end{figure*}
where $\beta_{i,\sigma_{i}}^{a}$ is the contribution of $p_{z}^{i}$
orbital of spin $\sigma_{i}$ to the single-electron eigenstate $a$,
and $H_{c}$ is the electron-electron interaction potential in the Coulomb form
\[
H_{C}=\frac{e^{2}}{4\pi\epsilon\epsilon_{0}r_{12}}
\]
with $r_{12}=|\boldsymbol{r_{1}}-\boldsymbol{r_{2}}|$. The dielectric
constant is taken as for the CNT coated in the  silicon dioxide 
$\epsilon=4$. We use the two-center approximation
to calculate the interaction matrix elements \cite{tc}.
For the on-site integral ($i=j$) we take $\langle p_{z}^{i}p_{z}^{j}|\frac{e^{2}}{4\pi\epsilon_{0}r_{ij}}|p_{z}^{i}p_{z}^{j}\rangle=16.522$
eV \cite{potasz} and for $i\neq j$ we use the formula
$\langle p_{z}^{i}p_{z}^{j}|\frac{1}{r_{ij}}|p_{z}^{k}p_{z}^{l}\rangle=\frac{1}{r_{ij}}\delta_{ik}\delta_{jl}$ \cite{Osambi}.

\begin{figure*}[htbp]
\includegraphics[scale=0.3]{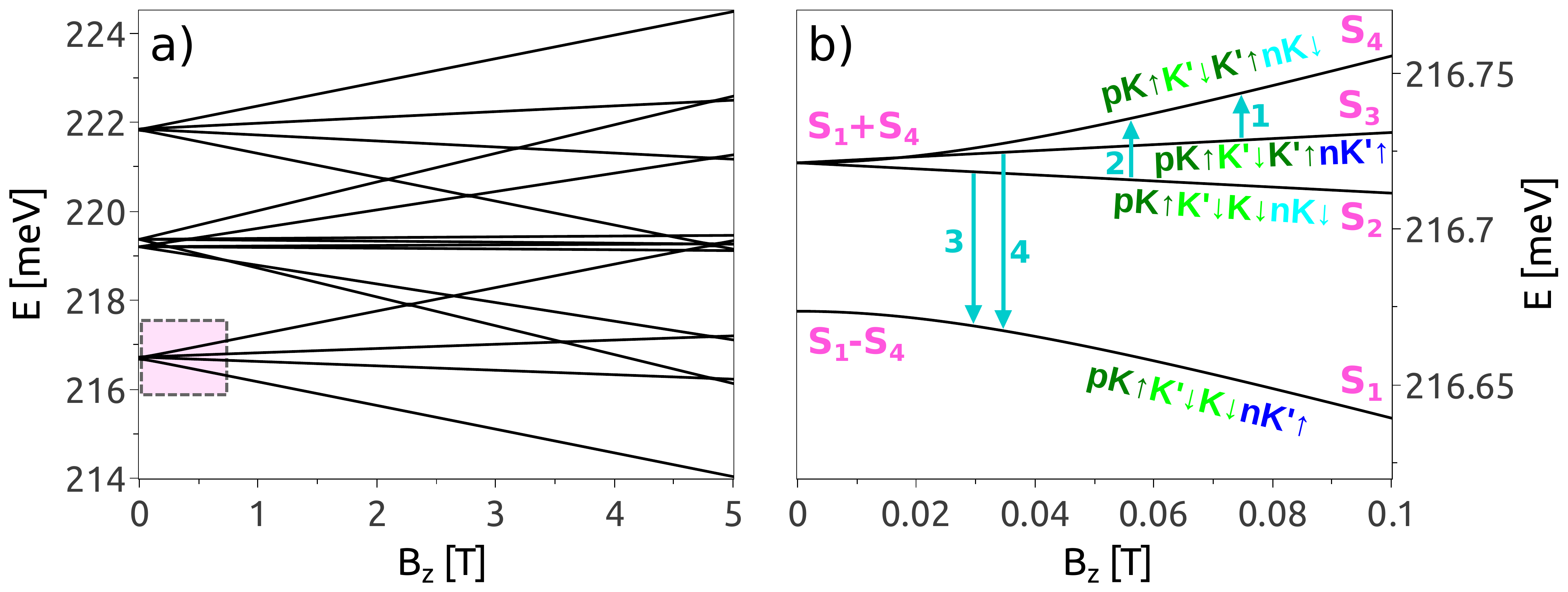}
\caption{ (a) The energy spectrum for 4 electrons for $V_{n}=220$ meV and $V_{p}=160$ meV.
All the plotted energy levels correspond to the (1e,1h) charge configuration.  
(b) Enlarged fragment of (a) with the 4 lowest (1e,1h) states. The labels placed near the energy levels indicate the {\it dominant} spin-valley configuration. The arrows (1-4) stand for the transitions considered in the paper.
Near the energy levels we indicate the spin-valley symmetry. $S_1,S_2,S_3$ and $S_4$ are the
wave functions of the subsystem of the last two-electrons: the highest-energy one in the $p$ dot and
the only one in the $n$ dot, that are discussed in the text.
} \label{qti}
\end{figure*}

\section{RESULTS}

\subsection{Straight CNTs}

\subsubsection{Single-carrier energy levels}
The single-electron spectrum is plotted in Fig. \ref{spectra}(a) as functions of $V=V_n=V_p$.
The energy levels plotted in blue (green) correspond to states localized within the $n$ ($p$) QD.
The energy levels of states localized within the $n$ ($p$) QD have positive (negative) energies that decrease (increase) with $V$.
For the simulation of the experimental workpoint with the (1e,1h) ground-state charge configuration 
we set the potential parameters to $V_n=220$ meV and $V_p=160$ meV [see the arrows in Fig. \ref{spectra}(a)].
The single-electron energy spectrum as a function of the magnetic field applied
along the axis of a straight CNT is given in Fig. \ref{spectra}(b).
The states of both the conduction and the valence bands correspond to Kramers doublets
which for $B=0$ are separated by the SO splitting $\Delta_{SOp}$ and $\Delta_{SOn}$  [see Fig. \ref{spectra}(b)].
The spectra for the energy levels localized in the $n$- and $p$-type QDs
display a similar dependence on the magnetic field, only the valley indices of the corresponding
energy levels are interchanged.

\subsubsection{Description of few-carrier states}
We deal with the single-electron states that appear near the charge neutrality point.
In the basis that spans Hamiltonian (\ref{4e}) we include 
the four highest-energy states of the valence band (plotted in light and dark shades of green in Fig. \ref{spectra}(b)
and marked by the lower arrow in Fig. \ref{spectra}(a)),
the four lowest-energy states of the conduction band (plotted in shades of blue in Fig. \ref{spectra}(b) and marked by the higher arrow
in Fig. \ref{spectra}(a)), as well as a number  of higher-energy conduction band states  (up to 16) which become partially occupied by the carriers in presence of the electron-electron interaction. The basis contains up to 24 single-electron spin-orbitals.
We assume that all the lower-energy valence band states are fully occupied.
The basis [Eq. (\ref{4e})] accounts for all the possible charge configurations of the last four carriers, and
 we focus our attention on the (1e,1h)  ground-state charge distribution 
[see the lower part of Fig. \ref{schemat}(a)],
 with 1 electron in the conduction-band-state localized in the $n$-dot and 3 electrons occupying states of the  top of the valence band localized
in the $p$-dot, leaving a single unoccupied level.    
The energy spectrum for the system of four electrons
is given in Fig. \ref{qti}(a). All the energy levels within this plot correspond to the (1e,1h) configuration. The other distributions appear
higher in the energy spectrum.
Figure \ref{qti}(b) displays the enlarged fragment of the spectrum  with a singlet ground-state
and a triplet excited state at $B=0$. 
 The axial magnetic field lifts the  degeneracy of the triplet.   In the figure we denote the occupied single-electron states that appear in the subsequent energy levels for $B=0.1$ T. 
For lower $B$ the states with zero spin projection on the $z$-axis: the ground state energy level and the third excited energy level enter into an avoided crossing.
In all the states of Fig. \ref{qti}(b),
the two lowest energy levels of the $p$-dot [p$K\uparrow$ and $pK'\downarrow$ in Fig. \ref{spectra}(b)]  are  occupied. For the remaining two-electrons: the electron in the $n$-type dot occupies one of the two lowest-energy states [n$K'\uparrow$ or n$K\downarrow$ -- see Fig. \ref{spectra}(b)],
and the electron in the $p$-type dot occupies one of the two highest-energy states [p$K\downarrow$ or  p$K'\uparrow$]. The ground-state multiplet of the four-electron system in the (1e,1h) configuration is thus effectively 
 spanned by a $K\downarrow$ and $K'\uparrow$ spin-valley states of the two last electrons -- one in the $n$-type dot 
and the other in the $p$-type dot. The basis  functions for these last two electrons can then be written as 
\begin{equation}S_1=\frac{1}{\sqrt{2}}\left(pK\downarrow(1) nK'\uparrow(2)- nK'\uparrow(1) pK\downarrow(2) \right),\end{equation}
\begin{equation}S_2=\frac{1}{\sqrt{2}}\left[p(1) n(2)- n(1) p(2) \right] K\downarrow(1)K\downarrow(2),\end{equation}
\begin{equation}S_3=\frac{1}{\sqrt{2}}\left[p(1) n(2)- n(1) p(2) \right] K'\uparrow(1)K'\uparrow(2), \end{equation}
and
\begin{equation}S_4=\frac{1}{\sqrt{2}}\left(pK'\uparrow(1) nK\downarrow(2)- nK\downarrow(1) pK'\uparrow(2) \right).\end{equation}
For $B>0.08$ T the Hamiltonian eigenstates can be identified with these wave functions [see Fig. \ref{qti}(b)].

The two central energy levels in Fig. \ref{qti}(b)  are spin-valley polarized. Thus they are not coupled by the Coulomb interaction to any other configuration in the ground-state quadruple,
hence the linear dependence on $B$.
The diagonal interaction matrix element for
state $S_2$ is
\begin{eqnarray*}\langle S_2 | H_c | S_2 \rangle &=&  \langle p(1)n(2)|H_c| p(1)n(2) \rangle \nonumber \\ &-& \langle p(1)n(2)|H_c| n(1)p(2) \rangle \nonumber \\&=&C-X,\end{eqnarray*} where $C$ is the Coulomb
integral and $X<0$ is the exchange integral. The same result is obtained for the other spin-valley polarized configuration $S_3$.

For the spin-unpolarized configurations $S_1$ and $S_4$ the off-diagonal interaction matrix elements are non-zero and equal  $\langle S_1 | H_c | S_4 \rangle=-X$
while the diagonal ones are $\langle S_1 | H_c | S_1 \rangle=\langle S_4 | H_c | S_4 \rangle=C$.
The Hamiltonian matrix diagonalized in the basis of the $S_1$ and $S_4$ mattrices produces eigenvalues $C+X$ for the ground state [$\Psi_{gs}=\frac{1}{\sqrt{2}} (S_1 -S_4)$]
and the excited energy level $C-X$ [$\Psi_{es}=\frac{1}{\sqrt{2}} (S_1+S_4)$] that at $B=0$ forms a triplet with the two spin-valley polarized states $S_2$ and $S_3$. 
The wave functions for the spin-valley unpolarized states at $B=0$ can be written as \begin{eqnarray}\Psi_{gs}&=& \frac{1}{\sqrt{2}}\left[p(1)n(2)+n(2)p(1)\right] \nonumber  \\ &\times &  \left[K\downarrow(1)K'\uparrow(2)-K'\uparrow(1)K\downarrow(2)\right]\end{eqnarray}
and \begin{eqnarray} \Psi_{es}&=& \frac{1}{\sqrt{2}}\left[p(1)n(2)-n(2)p(1)\right] \nonumber \\ &\times &\left[K\downarrow(1)K'\uparrow(2)+K'\uparrow(1)K\downarrow(2)\right].\label{later} \end{eqnarray}
In  Eq. (\ref{later}) the spatial wave function is antisymmetric with respect to the electrons interchange and forbids the electrons to 
occupy the same spatial orbital as it does for $S_2$ and $S_3$.

\begin{figure}[htbp]
\includegraphics[scale=0.26]{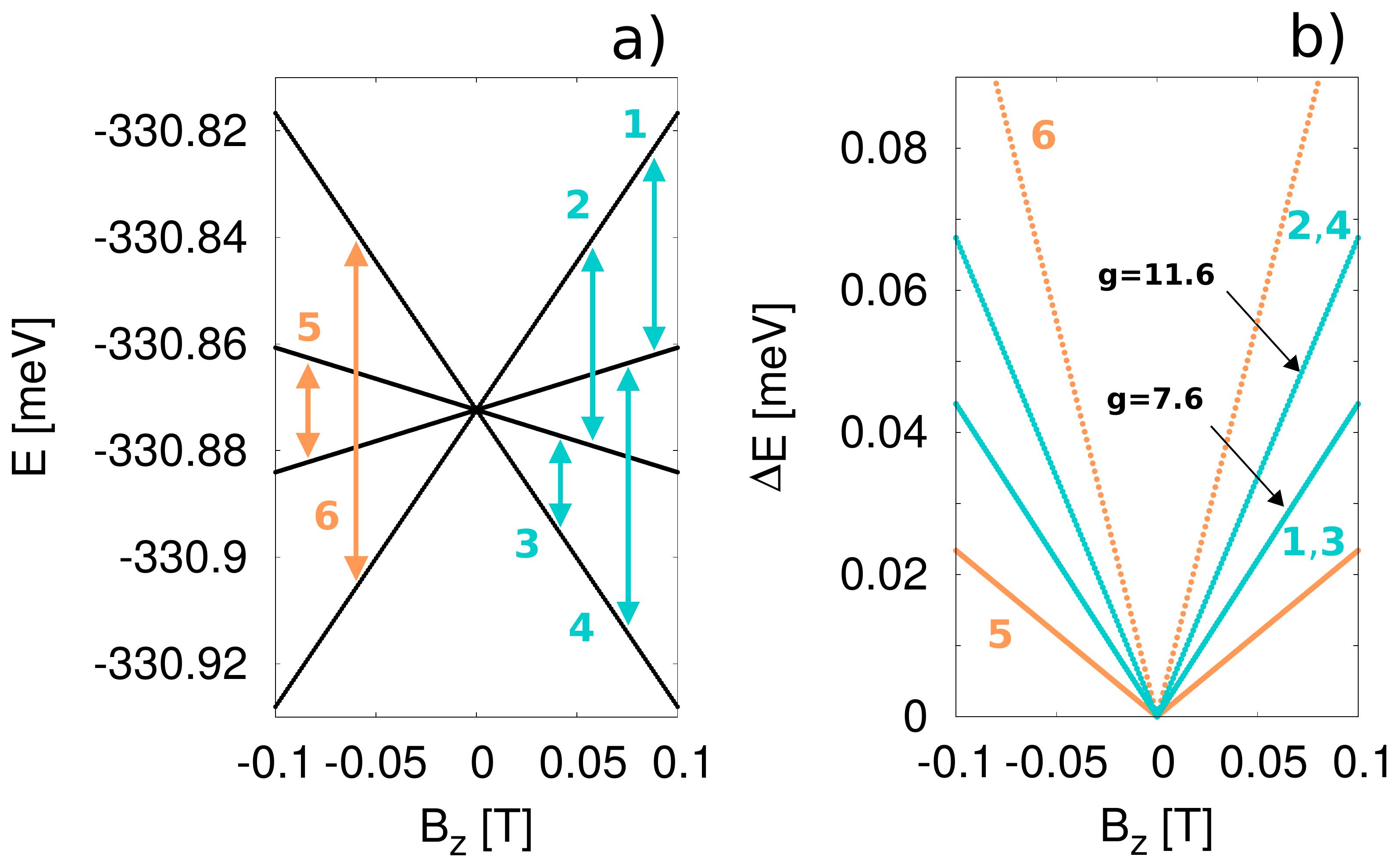}
\caption{(a) Energy spectrum as a function of the magnetic field parallel to the CNT axis ${\bf B}=(0,0,B_z)$ for the system of Fig. \ref{qti}(b) but without the electron-electron interaction. 
(b) Transitions lifting (blue) and conserving (orange) spin-valley blockade of the current -- indicated by arrows on (a). 
For the interacting case see Fig. \ref{prosta}(b) and  (e).} \label{wt}
\end{figure}

\begin{figure}[htbp]
\includegraphics[scale=0.26]{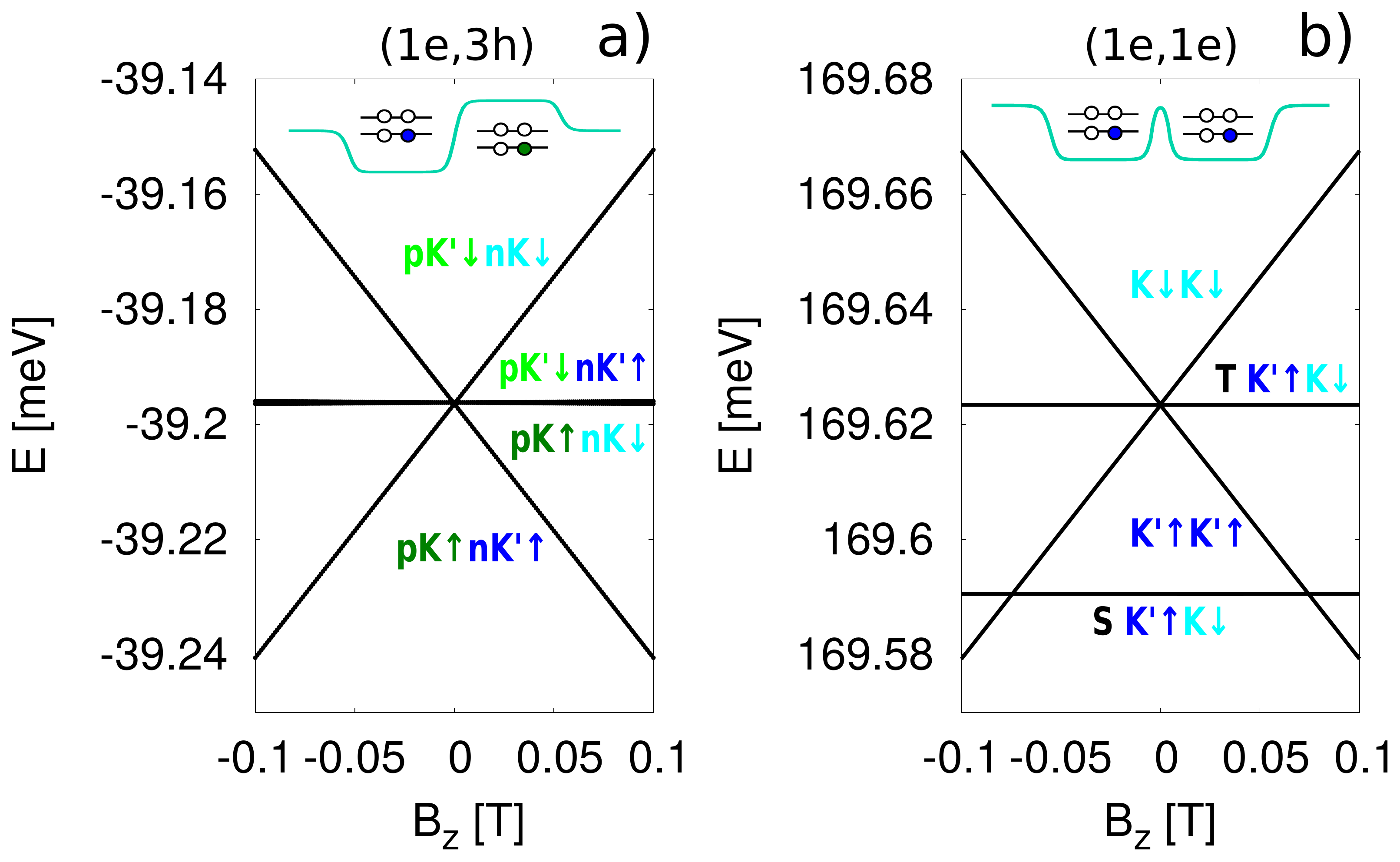}
\caption{Energy spectrum as a function of the magnetic field parallel to the CNT axis ${\bf B}=(0,0,B_z)$ for the system with a single electrons in $n$ and $p$-type dots -- for the charge configuration denoted as (1e,3h)  (a)  
and for the system with a single electron in two $n$-type dots -- (1e,1e) charge configuration (b). The electron-electron interaction
is included. In (a) no avoided crossing in the spectrum -- similar to the one of Fig. \ref{qti}(b) and Fig. \ref{prosta}(b) is obtained. The exchange interaction vanishes due to spin-valley orthogonality of
the single-dot states. In (b) the exchange interaction splits singlet-like (S) and triplet-like (T) states and remains constant for varying $B_z$. } \label{1e3h}
\end{figure}

Concluding the analysis at $B=0$, we find  a singlet ground state and an  excited energy level which is three-fold degenerate 
-- as in the system of two-electrons in a double QD defined within the III-V semiconductors \cite{exchange}. For comparison, the spectrum in the absence of the electron-electron interaction
is given in Fig. \ref{wt}(a) which indicates a fourfold degenerate ground-state.
The analogy of the ground-state singlet-triplet structure of the (1e,1h) charge configuration with III-V QDs \cite{exchange} 
 or with CNT unipolar n-n double QDs ends at zero-magnetic field.
For (1e,1e) states, in III-V's \cite{exchange} as well as in CNTs [Fig. \ref{1e3h}(b)] in the external magnetic field a spin-polarized state replaces the singlet in the ground-state.   
No singlet-triplet crossing is observed in the ground-state of the (1e,1h) system -- see Fig. \ref{qti}(b).  
The magnetic field shifts down the $S_1$ configuration on the energy scale faster than the rest of the configurations of Fig. \ref{qti}(b).
The $S_1$ is the only configuration in which both the last two carriers occupy the  energy levels that decrease in the external magnetic field [Fig. \ref{spectra}(b)]:
$pK\downarrow$ and $nK'\uparrow$. On the other hand in $S_4$ the energy levels of the last two carriers increase in $B$.
For this reason the  magnetic field lifts the mixing of the $S_1$ and $S_4$ states, and 
the ground-state at $B\simeq 0.1$ T corresponds to a pure $S_1$ state.

The structure of the spectrum of the (1e,1h) system is also in a striking  contrast to the (1e,3h) configuration [Fig. \ref{1e3h}(a)] with a fourfold 
degenerate ground state at $B=0$. In (1e,3h) configuration [see the inset to Fig.\ref{1e3h}(a)] the electron in the $n$ dot
occupies $K\downarrow$ or $K'\uparrow$ state, and the electron in the $p$ dot:  $K'\downarrow$ or $K\uparrow$.
The exchange integrals vanish for all occupations due to the spin-valley orthogonality  hence the ground-state degeneracy at $B=0$.

\subsubsection{Transitions spectra}

In the experiment \cite{lairdpei} the electron flow involves the charge hopping in a sequence (1e,1h) $\rightarrow$ (0e,0h).
The flow gets blocked in (1e,1h) configuration when the electron in the $n$ dot happens to occupy an energy level corresponding to a spin-valley
state that is already occupied by one of the three electrons of the $p$ dot [see Fig. 1(a)].
In terms of the few-particle states the spin-valley blockade occurs for the two central energy levels $S_2$ and $S_3$
of Fig. \ref{qti}(b).
The flow gets unblocked when the system is  transferred to the lowest ($S_1$) or the highest ($S_4$) energy level of Fig. \ref{qti}(b).
The experimental data are given in Fig. S9 of Ref. \cite{peilaird} and in Fig. 2 of Ref. \cite{lairdpei}. 
The transitions that lift the blockade are marked by the arrows in Fig \ref{qti}(b). 

In the absence of the magnetic field  $\Psi_{es}$ given by Eq. (\ref{later}) has
an antisymmetric spatial wave function. The Pauli blockade can be expected to appear in this state (see Ref. \cite{lilaird}) as it does for $S_2$ and $S_3$, and similarly in $T_0$ state for the electron pair in III-V QDs \cite{exchange}. 
For this blockade to occur in $\Psi_{es}$ state, the $S_1$ and $S_4$ states need to contribute {\it evenly} which only occurs at $B \simeq 0$ [see Fig. \ref{qti}(b)].
We find that for $B=0.02$ T, 0.05T and 0.1 T the contributions of the $S_1$, and $S_4$ to the ground state 
are 0.63, 0.26 ; 0.78, 0.11; and 0.86, 0.04, respectively. For the excited unpolarized energy level these contributions are reversed.
In the magnetic field  the ground state tends to $S_1$ and the excited state tends to $S_4$, which are not Pauli blocked.
Also, any other effect lifting the valley degeneracy -- crystal defects for instance -- will unblock the $\Psi_{es}$ state also at $B=0$. The lifted Pauli blockade for $\Psi_{es}$ is evident
in the experimental data (cf. Fig. 2 of Ref. \cite{lairdpei}) by the mere presence of the transition lines that reach a near-zero frequency at $B=0$. These experimental lines can only 
correspond to EDSR transitions from $S_2$ and $S_3$ to $\Phi_{es}$.

\begin{figure}[htbp]
\includegraphics[scale=0.22]{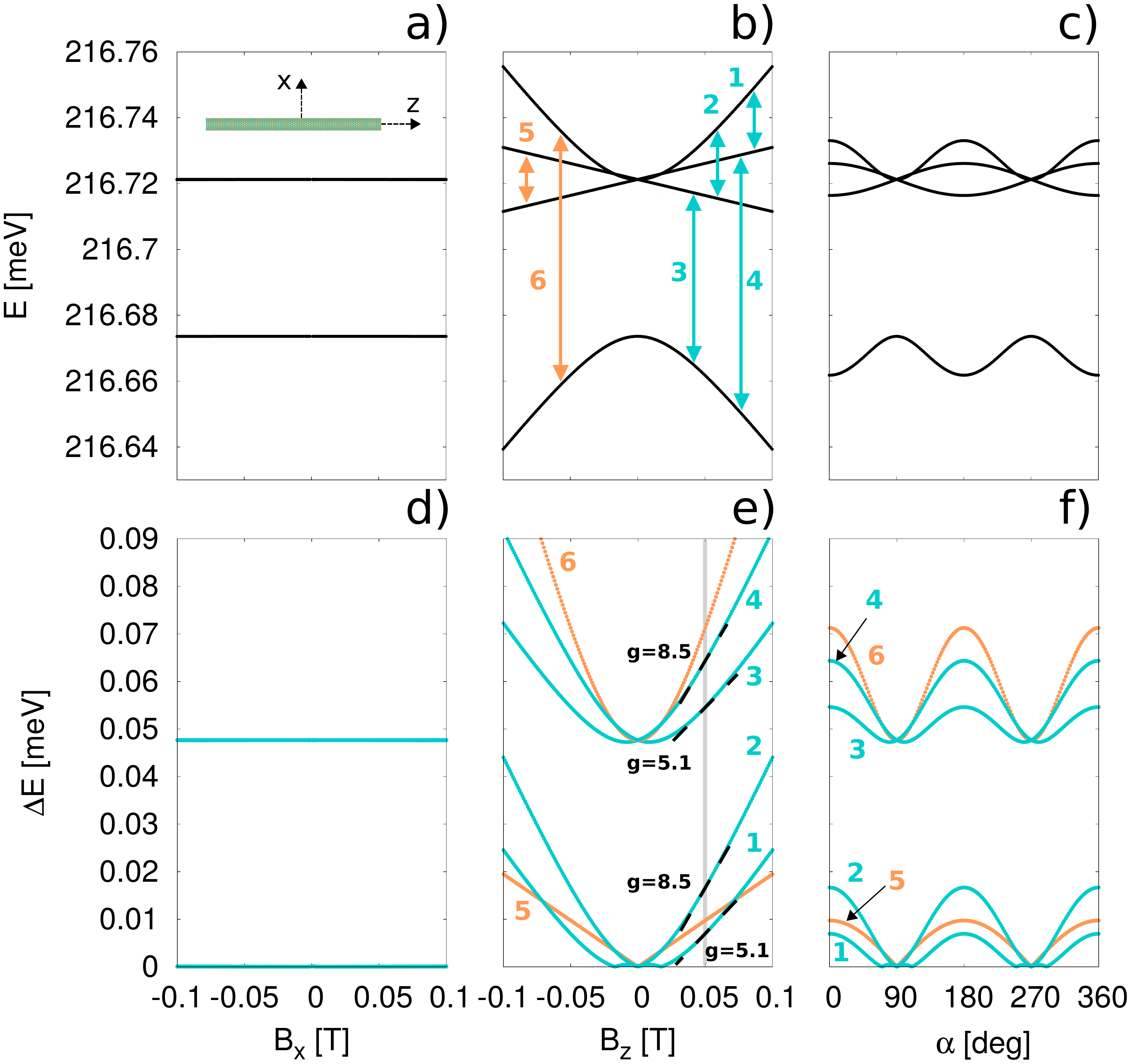}
\caption{Energy spectrum of the four electron system for a straight CNT with parameters of Fig. \ref{qti}, for the magnetic field ${\bf B}=(B_x,0,0)$ perpendicular
to the axis of the tube $z$ (a), for the magnetic field parallel to the CNT axis ${\bf B}=(0,0,B_z)$ (b), and as a function of the angle $\alpha$ [see Fig. \ref{schemat}(a)]
between the magnetic field vector and the $z$ axis for $B=0.05$ T, ${\bf B}=B(\sin\alpha,0,\cos\alpha)$. (d,e,f) -- Same as (a,b,c) only for the transition energies
that are defined in (b). In blue we plotted the transitions between spin-valley polarized and unpolarized energy levels -  numbered as in Fig. \ref{qti}(b) by (1-4).
These transitions involve lifting of the spin-valley blockade of the current. The transitions (5) and (6) occur between spin-valley polarized (unpolarized) states.}\label{prosta}
\end{figure}

The transition energies from the blocked states to the unblocked ones 
are plotted in blue in Fig. \ref{prosta}(e) for a straight CNT and an axial magnetic field.  These transitions are also marked in the energy spectrum in Fig. \ref{prosta}(b).
The weak dependence of the energy levels on the magnetic field perpendicular to the CNT axis  [see Fig. \ref{prosta}(a) and \ref{spectra}(b)] is a consequence of the SO coupling, which polarizes
the spins in the direction parallel to the CNT axis. 
Energy change of the energy levels in Fig. \ref{prosta}(a) is of the order of $10^{-3}$ meV/T.
When the magnetic field is rotated within the $xz$ plane [Fig. \ref{prosta}(c,f)] the transition lines
change due to the Zeeman effect corresponding to the $z$ component of the magnetic field $B_z=B \cos \alpha$ (in Fig. \ref{prosta}(c,f) $B=0.05$ T).

The transition energies plotted in Fig. \ref{prosta}(e) for $B_z\simeq 0$ -- within the range of the avoided crossing opened between $S_1$ and $S_4$, 
  are not linear functions of the magnetic field. 
 The effective $g$-factors
for these transitions ($g=\frac{1}{\mu_B} \frac{dE}{dB_z}$) as extracted for $B_z=0.05$T, off the avoided crossing, i.e., in the linear part of the dependence, are given in
the Figure. The calculated effective $g$ factors are: $g_{1,3}=5.1$ for transitions 1 and 3, and $g_{2,4}=8.5$ for transitions 2 and 4.
As we show below also for bent CNTs the effective Land\'e factors keep always  equal values of the $g$ factors for the pairs of transitions (2,4) and (1,3).
 The main source of this splitting is the spin Zeeman interaction for spin polarization along the axis of the tube that results from the SO coupling. The spin Zeeman effect 
alone amounts in a difference $\Delta g=g_{2,4}-g_{1,3}=4$ -- the result that we exactly reproduce for neglected electron-electron interaction -- see Fig. \ref{wt}(b).
In presence of the electron-electron interaction for various bends we find $\Delta g \in (3.2,3.8)$.  

The experimental data for the  EDSR frequencies lifting  the current blockade as given in Fig. 2(c) of Ref. \cite{lairdpei}
 agree with the results plotted in blue in  Fig. \ref{prosta}(f) in two aspects: {\it (i)}  two pairs
of transitions shifted on the energy scale are observed, and {\it (ii)}  the transitions reach maxima for $\alpha=0$ and $\pi$ and minima for $\alpha=\pi/2$ and $3\pi/2$.
 In the experimental data the lower-energy transition
acquires non-zero values at minima indicating a splitting of the triplet degeneracy at $B=0$ by e.g. a residual hyperfine interaction with $^{13}C$ atoms.
 In the theoretical results  [Fig. \ref{prosta}(f)] the splitting is the same for both the lower and the upper branches and goes to zero for the magnetic field
oriented perpendicular to the axis of the CNT $\alpha=\pi/2$ and $3\pi/2$. In the experiment  \cite{lairdpei} the splitting of the upper branch is distinctly larger and the splittings pertain for 
$B$ oriented along the $x$ direction, which again points to a lifted triplet degeneracy at $B=0$.

The experimental data as a function of the axial magnetic field (Fig. 2(b) of Ref. \cite{lairdpei}) indicates a nearly 
linear transition spectrum at $B \simeq 0$. The curvature of the calculated results of Fig. \ref{prosta}(e) is due to the $S_1$--$S_4$ avoided crossing.
Obviously the nonlinear  part of the spectrum appears within a narrower region of the magnetic fields
for weaker tunnel coupling between the dots (not shown). 

The experimental \cite{lairdpei} effective $g$ factors are roughly  twice smaller than the ones calculated in the present model. 
For axial orientation of the magnetic field the $g$ factors are dominated by the orbital effects, and their reduction is expected for a smaller diameter of the CNT (see Eq. (10) of Ref. \cite{rmp}). 
The experimental $g$ factors for the  pair of transition lines in the upper branch of the spectrum differ by 1.5 only which is far from 4 -- the value 
expected from the spin Zeeman factor. Moreover, in the experiment no difference in the $g$ factors is observed in the lower branch of energy levels. 
In fact the results of Fig. \ref{prosta}(e) agree well with Fig. S9 of Ref. \cite{peilaird} -- with the parabolic dependence
of  the transition lines and the shifts of the minima to off zero magnetic field and the splitting of the pairs of transition lines. 
However, Fig. S9 of Ref. \cite{peilaird} corresponds to the
nominally perpendicular orientation of the magnetic field. The other result for this orientation of the magnetic field -- of Fig. 2(a) of Ref. \cite{lairdpei} indicates a 
$g$ factors of the order of 2 and coalescence of the transition lines for $B_z\geq 50$ mT. The present results -- for a straight CNT -- Fig. \ref{prosta}(a) 
indicate no dependence on $B_x$ in the considered range. In the next subsection we make an attempt to account for the features observed
in the experiment by a bend of the CNT.

\subsection{Bent CNTs}

\begin{figure*}[htbp]
\includegraphics[scale=0.34]{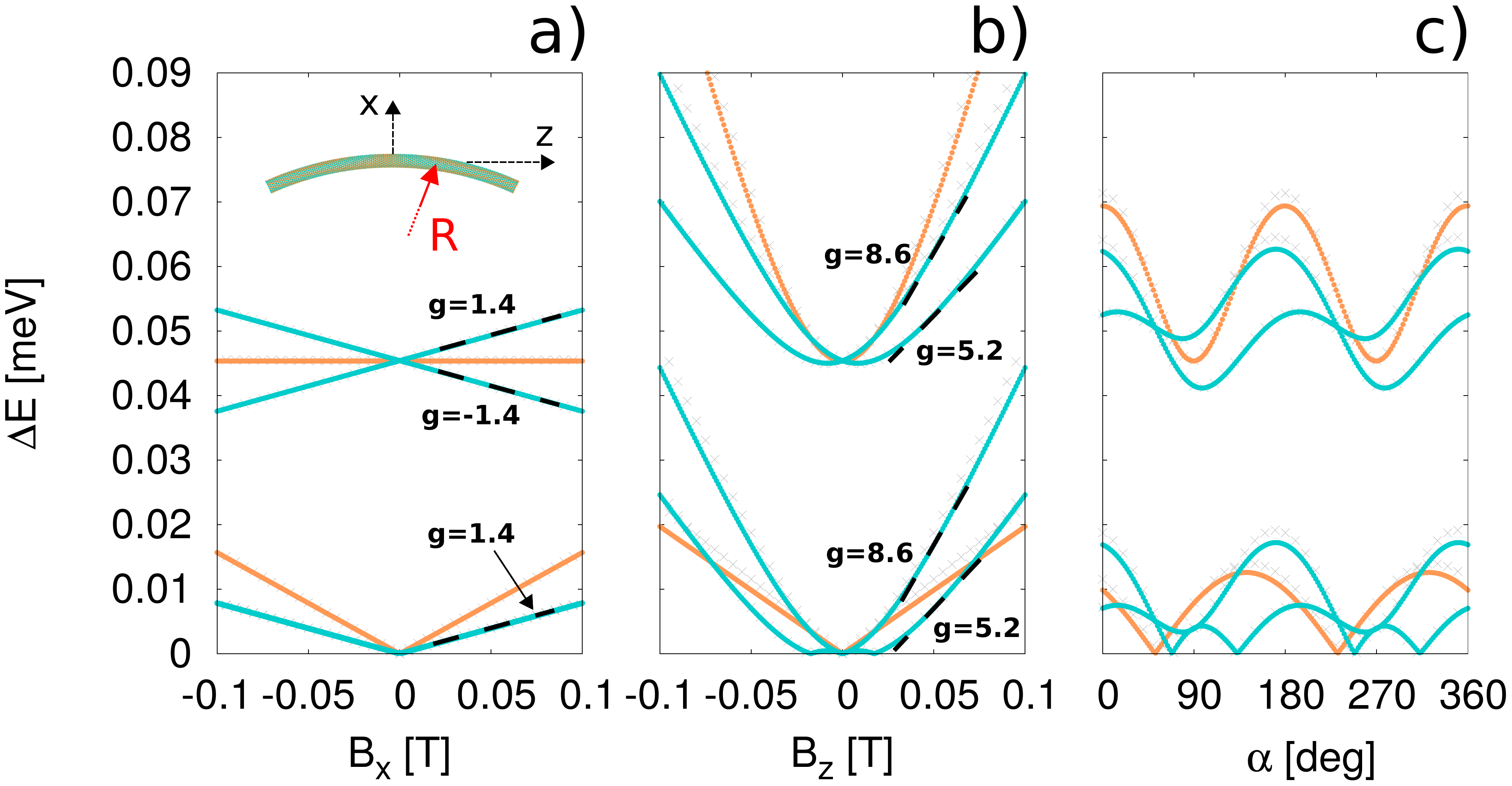}
\caption{Transition energies as a function of $B_x$ (a) and $B_z$  (b). Plot (c) shows the angular dependence on the orientation
of the magnetic field vector with respect to the $z$ axis. The CNT is bent to form an arc of radius $R=60$ nm - compare with the results for a straight CNT displayed in Fig. \ref{prosta}(d-f).
Grey crosses correspond to the results obtained with the simple model described by the Hamiltonian (\ref{eq:4x4}). 
} \label{60nm}
\end{figure*}

Figure \ref{60nm} displays the transition energies for a CNT bent into an arc of radius 60 nm (see inset)
for the magnetic field oriented parallel to the $x$ direction [Fig. \ref{60nm}(a)], to the $z$ direction [Fig. \ref{60nm}(b)],
and as a function of the angle  formed by the {\bf B} vector with  the $z$ axis for $B=0.05$ T [Fig. \ref{60nm}(c)]. 
The energy levels $S_2$ and $S_3$ 
split in nonzero
$B_{x}$.
 We find that the states with 2 electrons in $K$ and 2 in $K'$ valleys are insensitive
to the modulation of $B_x$. On the other hand, the states with 3 electrons in one of the valleys
are split by $B_x$.

\begin{figure*}[htbp]
\includegraphics[scale=0.34]{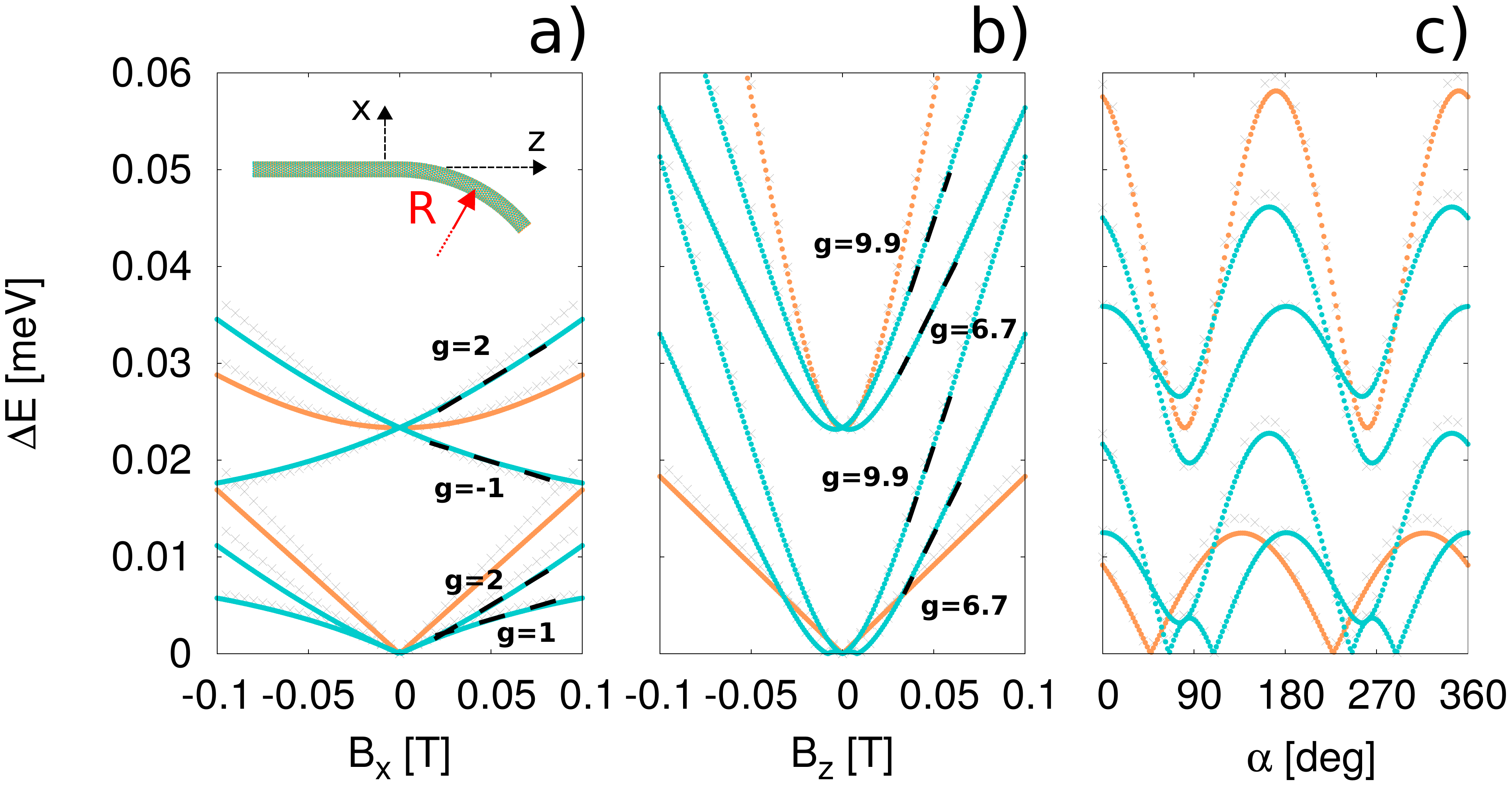}
\caption{Transition energies as a function of $B_x$ (a), $B_z$ (b) and $\alpha$ (c). Nanotube bent into an arc of a bending radius $R=30$
nm on the $p$-dot side and straight on the $n$-dot side.}\label{wp}
\end{figure*}

The dependence of the transition energies  on $B_z$ is only weakly affected by the bend.
However, the transition energies do react to the field in the $x$ direction, which makes the calculated results qualitatively closer to the experiment. 
The bend also destroys the symmetry of the transition plot as a function of the orientation angle [Fig. \ref{60nm}(c)].
For the transition spectrum distorted by the bend, the transition energies for the magnetic field oriented along $z$ or $-z$ direction ($\alpha=0$ or $\alpha=\pi$) are still identical, and similarly for $x$ and $-x$ directions  ($\alpha=\pi/2$ and $3\pi/2$), but the transition lines are no longer symmetric
with respect to these angles, and  no extrema of the calculated transition energies are found at these angles.
In the experimental data of Fig. 2(c) of Ref. \cite{lairdpei} transition lines are symmetrical with respect to the angles $\alpha=\pi/2$ and $3\pi/2$, which resembles rather the previous result obtained for a straight CNT [Fig. \ref{prosta}(c)].
However, experimental data is not symmetrical with respect to the angles $\alpha=0$ or $\alpha=\pi$ -- this time in agreement with the spectra of Fig. \ref{60nm}(c).

For the CNT bent  only at  the  $p$ side  (Fig. \ref{wp}, for arc radius of 30 nm) all the energy levels react to the $B_x$ field. This is because now $B_x$ interacts only
with 3 electrons in the $p$ dot, so one never has an equal number of electrons in both the valleys.
The $g$ factors for the $B_z$ dependence are increased with respect to the precedent case. 
The increase of $g$ factors in Fig. \ref{wp} with respect to Fig. \ref{60nm} results from a decreased tunnel coupling between the dots, which makes the avoided
crossing between energy levels $S_1$ and $S_4$ thinner. Moreover, the $g$ factors for the $B_x$ orientation are no longer identical in their absolute values Fig. \ref{wp}(a), 
which is also the case for the experimental data of Fig. 2(a) of Ref. \cite{lairdpei}.
The angular dependence of the transition energies Fig. \ref{wp}(c) preserves the form of Fig. \ref{60nm}(c), i.e. it deviates
qualitatively from the experimental data.

\begin{widetext}
We have found that the transition energies spectra of Figs. \ref{60nm} and \ref{wp} can be quite accurately 
described in a simple model which maps the bend of the CNT to the rotation of the effective magnetic field felt by the carriers in 
each of the dots. 
For this purpose we took a Hamiltonian written in the basis of $S_i$ wave functions [Eq. (5-8)], 
\begin{equation}\label{eq:4x4}
H_s=\left(\begin{array}{cccc} C+E_B(S_1)&0&0&-X\\ 0&C-X+E_B(S_2)&0&0 \\ 0&0&C-X+E_B(S_3)&0 \\ -X&0&0&C+E_B(S_4) \end{array}\right),
\end{equation}
where $C$ and $X$ are the Coulomb and exchange energies defined above.
The terms $E_B(S_i)$ account for the shifts of the single-electron energy levels in the external magnetic field.
 For each of the basis elements the terms are calculated according to $E_B(S_i)=\sum_{j=1}^2 (E_j + g_{||}^j\mu_B B'^j_z) $
where the summation over $j$ runs over the single-electron states occupied in the $S_i$ configuration, $E_j$ is $j$th single electron energy for $B=0$, 
$g_{||}^j$ is the effective $g$ factor calculated for a straight CNT and an axial magnetic field (Fig. \ref{spectra}(b)), and $B'^j_z=\langle \Psi_j |{ {\bf B} \cdot {\bf z(\phi)}}|\Psi_j\rangle $, where $\Psi_j$ is the eigenfunction for the bent CNT and ${\bf z}(\phi)=(-\sin\phi,0,\cos\phi)$ is the versor of the local CNT axis [see Fig. 1(c)]. 
The results for the transition energies calculated with this model [crosses in Fig. 7 and 8] are in a good agreement with the complete numerical scheme.
\end{widetext}

\begin{figure}
\includegraphics[scale=0.22]{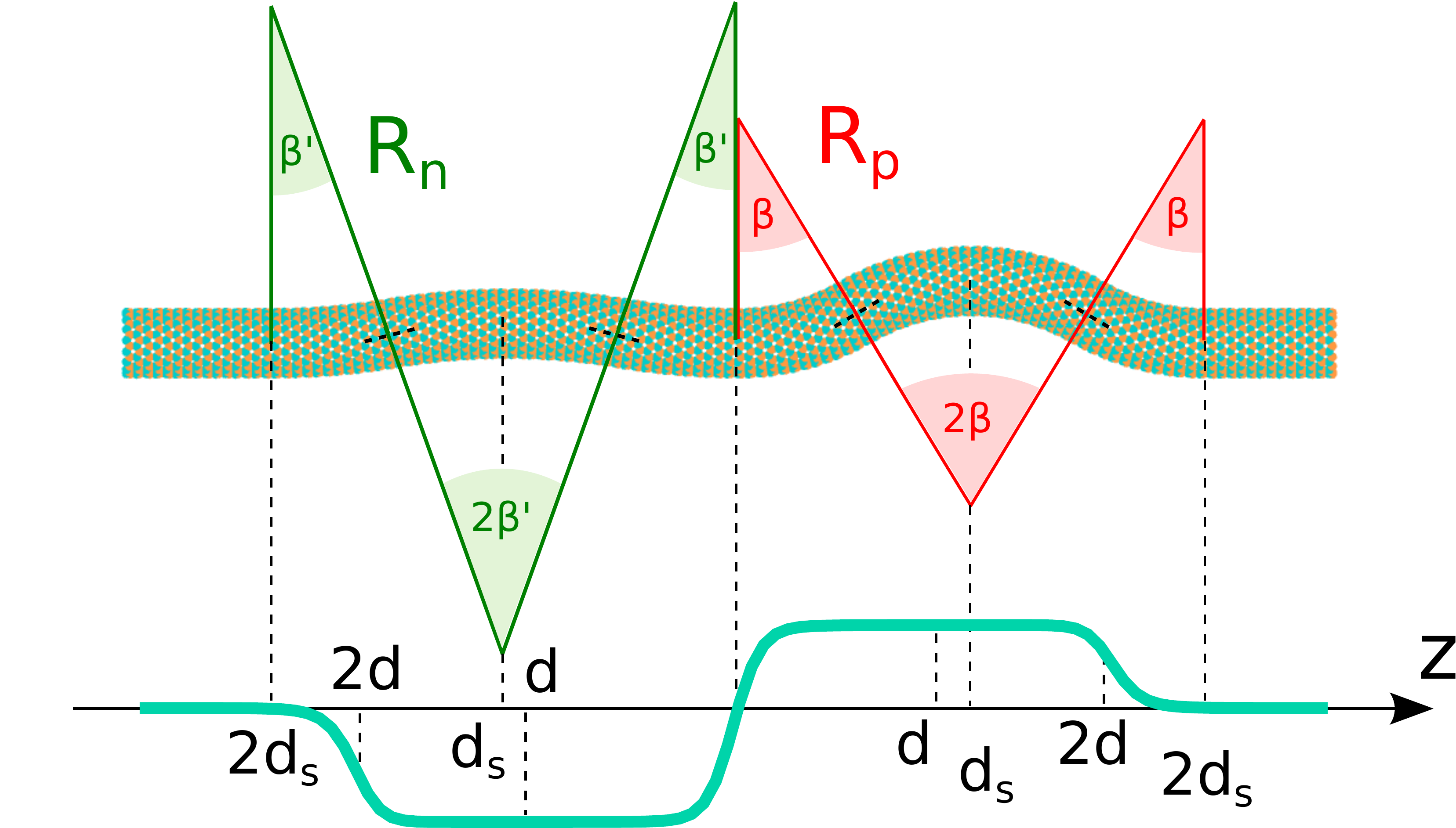}
\caption{Schematics of the considered system with CNT bent above the metalic gates.} \label{wa}
\end{figure}

\begin{figure}
\includegraphics[scale=0.22]{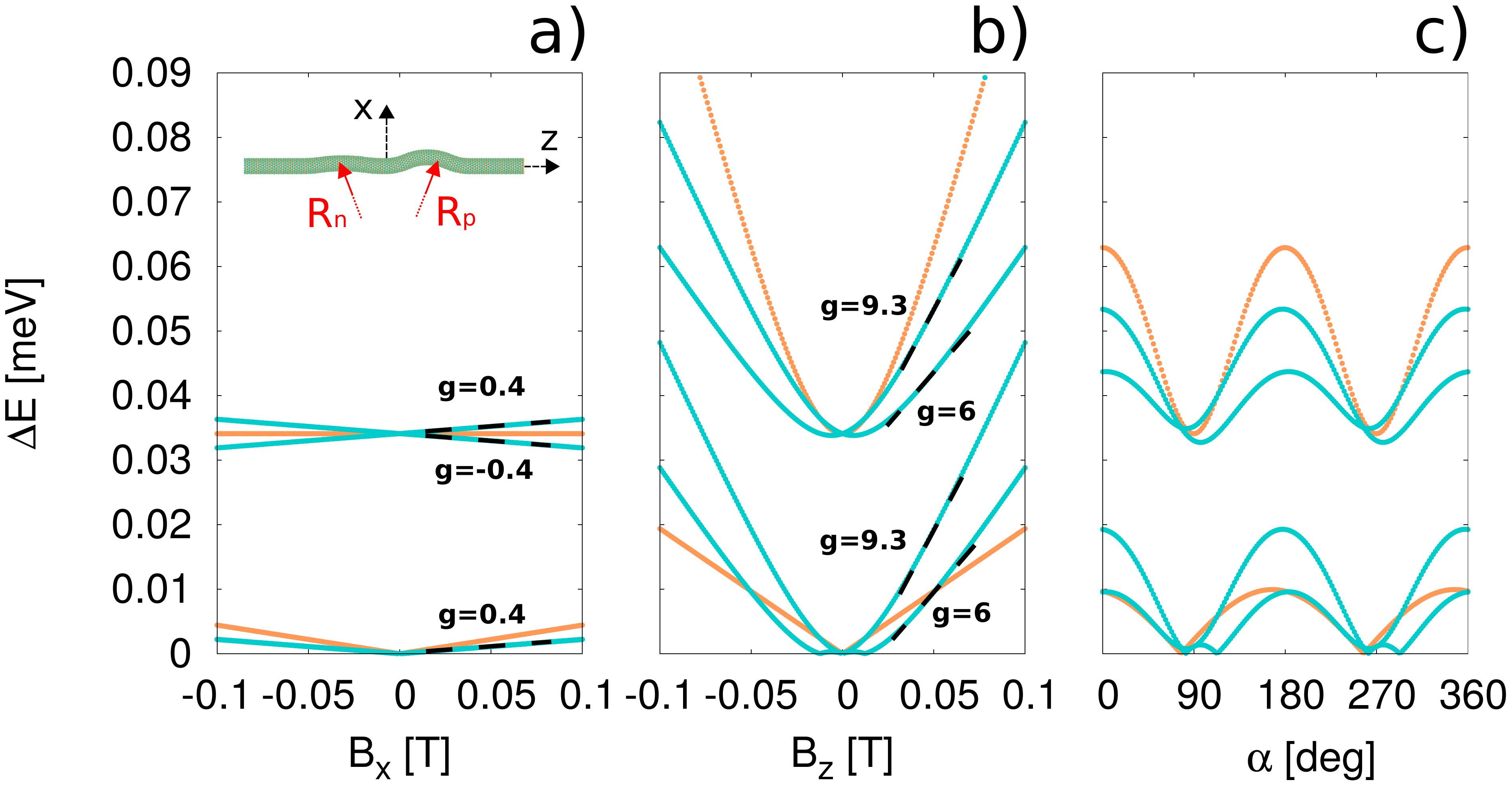}
\caption{Transition energies as a function of $B_x$ (a), $B_z$ (b) and $\alpha$ (c). Nanotube bent above the metalic gates as of Fig. \ref{wa} with bending radius on the $n$
side $R_{n}=30$ nm and $p$ side $R_{p}=10$ nm and $d=d_s=8$ nm.} \label{wat}
\end{figure}




\begin{figure*}
\includegraphics[scale=0.35]{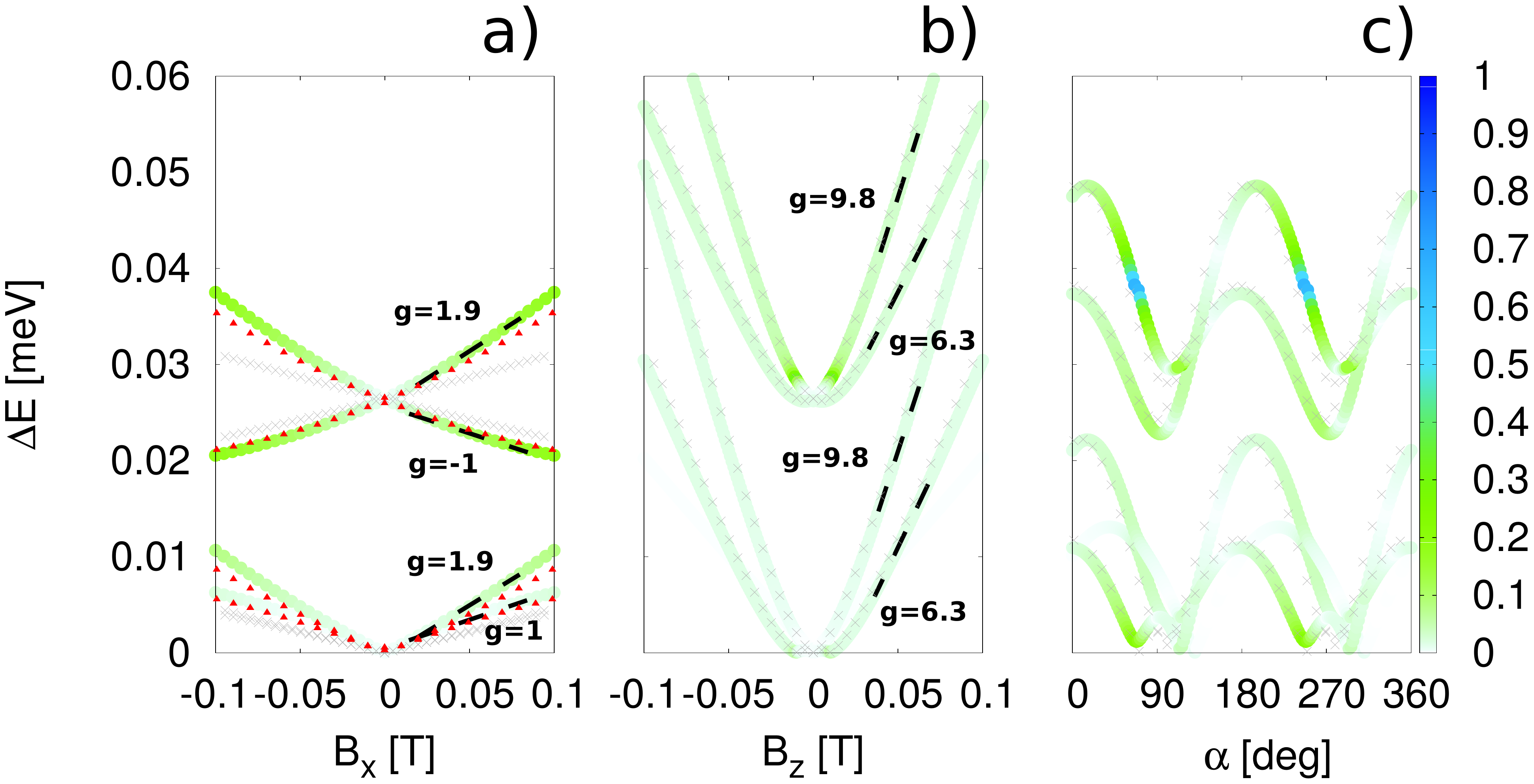}
\caption{Results for schematics of Fig. \ref{wa} with $R_{n}=30$ nm, $R_{p}=10$ nm and $d=8$ nm, $d_s=10$ nm. The lines indicate the intensities of the transitions
that lift the spin-valley blockade: labelled as (1,2,3) and (4) in Fig. \ref{qti}(b) and Fig. \ref{prosta}(b,e,f). 
The color of the points indicates the magnitude of the matrix element $|\langle \phi_i|\sum_ j z_j|\phi_f\rangle|$, where
$z_j$ corresponds to the amplidute of the ac electric field, summation $\sum_ j$ is over 4 electrons, $\phi_i$ and $\phi_f$ corresponds to the initial 
and final states. Grey crosses correspond to the transision spectra obtained with the simple model described by the Hamiltonian (\ref{eq:4x4}). Red triangles stand for results not accounting for
the SO coupling due to the bend of CNT axis.
} \label{wi}
\end{figure*}

The best agreement with the experiment is obtained when the bend of the CNTs is correlated with the variation of the potential profile
-- which corresponds to the metal gates deforming the CNT -- see Fig. 1(c,d) of Ref. \cite{lairdpei}. 
The schematics of the simulated setup is given in Fig. \ref{wa}. The radii of the arc were taken unequal for the $n$ and $p$ side $R_n=30$ nm, and $R_p=10$ nm -- since accidental symmetry in this point is rather unlikely.
In Fig. \ref{wat} the transition lines are given for the bends matched exactly to the width of QDs ($d=d_s$).
In Fig. \ref{wi} we introduce a small misalignment of the bends and QDs centers ($d=8$ nm, $d_s=10$ nm). The main effect of the modification is a very pronounced increase of $g$ factors for $B_x$.
We added here an additional  peak of the potential (1eV) on a single atomic site at $z=-13.6$ nm which behaves as an intervalley scattering center coupling.
The quantitative effect of the peak on the energy spectra of Fig. \ref{wi} is negligible.  However, the intervalley coupling it introduces is crucial for the transition matrix elements.
Without the intervalley scattering the matrix elements for the transitions between the initial (spin-valley polarized) and final (spin-valley unpolarized) states driven by the ac field applied in the $z$ direction $\langle \phi_i|\sum_ j z_j|\phi_f\rangle$ are nearly zero and the defect increases the corresponding values
by a few orders of magnitude. 
The lines in Fig. \ref{wi} are  plotted in color which denotes the absolute value of the  transition matrix elements.  The results for the angular dependence [Fig. \ref{wi}(c)] indicate a maximal intensity for the maximal derivative of the transition energy over the magnetic field orientation angle
$-\frac{d}{d\alpha}\Delta E$, where anticrossing involving $S_1$ and $S_4$ takes place.  Only small shifts of extrema on the plot off multiples of $\pi/2$ are observed. 
In the $B_x$ dependence one of the lines in the low-energy branch is distinctly stronger than the other. In the upper branch
both the lines are of a comparable intensity and their crossing is well visible.

The crosses in Fig. \ref{wi} indicate the results calculated with the approximate model of Eq. (11). Here, the variation of the energy levels in $B_x$ is
underestimated, since the projection of the $x$ versor on the local CNT axis changes sign near the center of the QD, so that the averaged effective magnetic field
is small. In Fig. 11(a) with red tringles we also plotted the energy spectrum calculated for the SO coupling only due 
to the folding of the graphene plane into the nanotube \cite{Ando, osimre} -- without the correction to the SO coupling due to the bend of the CNT axis  \cite{OsikaJPCM}. The energy variation due to the bend-related SO contribution is small, however the EDSR spin-flipping transition rates can be increased several times \cite{OsikaJPCM}.





\section{Summary and conclusions}

We studied the (1e,1h) charge configuration in a $n$-$p$ double QD defined by external gates in 
a semiconducting carbon nanotube. The study was performed using an atomistic tight-binding approach with inclusion of the spin-orbit coupling that determines the spin-valley symmetries of the systems near the charge neutrality point. The systems of a few confined carriers were
studied with the configuration interaction method. 
We discussed the symmetries of the systems with respect to (1e,3h) and (1e,1e) charge distributions. In contrast to (1e,3h), the (1e,1h) charge configuration
has a non-degenerate ground-state due to a non-vanishing exchange energy. In contrast to (1e,1e) system, for (1e,1h) no crossing of energy levels in the ground-state is found. Moreover, in the excited
triplet of the (1e,1h) system the spin-unpolarized state has a spatially antisymmetric wave function
only for $B=0$, and for $B>0$ the spin-valley and spatial wave functions separately do not possess a 
well-defined symmetry with respect to the electrons interchange.

We have discussed the transitions lifting the spin-valley Pauli blockade in the double dot,
identifying two pairs of transition lines split at $B=0$ 
by an energy gap of the order of 0.1 meV in agreement with the experimental data.
The energy gap results from the avoided crossing between the spin-unpolarized ground-state 
and the spin-unpolarized excited state belonging to the triplet at $B=0$. We discussed the $g$ factors
characterizing separate transitions as well as the dependence of the transition energies 
on the orientation of the external magnetic field. 
We demonstrated that the bend of the CNT has a weak influence on the transition spectra for the
magnetic field oriented along the nominal axis of the tube. However, its effects for the other orientations of the magnetic field are pronounced. The performed study indicates that the calculated transition spectra agree best with the experimental data for CNTs  locally bent  within the gated region. 

\section*{Acknowledgments}
This work was supported by the National Science Centre
according to decision DEC-2013/11/B/ST3/03837. 
E.N.O. benefits from the doctoral stipend ETIUDA of the National Science Centre
according to decision DEC-2015/16/T/ST3/00266.
Calculations were performed in the PL-Grid Infrastructure.
E.N.O. acknowledges support from the Spanish MINECO (Severo Ochoa Grant No. SEV-2015-0522 and FOQUS FIS2013-46768) and the Generalitat de Catalunya (SGR 874).


\begin{thebibliography}{00}
\bibitem{weaker} A weaker tunnel coupling for the ambipolar charge configuration of the double dot is evident in Fig. 1c of Ref. \cite{peilaird}.
\end{thebibliography}
\end{document}